\documentclass[english,12pt]{article}
          \usepackage{babel}
          \usepackage[T1]{fontenc}
          \usepackage[utf8]{inputenc}
          \usepackage{pslatex}
          \usepackage[export]{adjustbox}
          \usepackage[dvipsnames]{xcolor}
          \usepackage{amsmath}
          \usepackage{amssymb}
          \usepackage{longtable}
          \usepackage{placeins}
\begin{document}
\begin{flushright}
\textit{ACTA ASTRONOMICA}\\
Vol.0 (2025) pp.1-28
\end{flushright}
\bigskip

\begin{center}
{\bf Superhumps in active dwarf novae. \\ Part I: ER Ursae Majoris}\\
\vspace{0.8cm}
{K. ~B~\k{a}~k~o~w~s~k~a$^{1}$, B. ~K~i~r~p~l~u~k$^{1}$, P. ~Z~i~e~l~i~\'{n}~s~k~i$^{1}$, M. ~M~o~t~y~l~i~\'{n}~s~k~i$^{1}$,\\ 
A. ~G~u~r~g~u~l$^{1}$, J. ~G~o~l~o~n~k~a$^{1}$, K. ~S~z~y~s~z~k~a$^{1}$}\\
\vspace{0.5cm}
\begin{small}
{$^1$ Institute of Astronomy, Faculty of Physics, Astronomy and Informatics,\\
Nicolaus Copernicus University, ul. Grudzi\k{a}dzka 5,
87-100 Toru\'{n}, Poland}\\
\end{small}
{\tt e-mail: bakowska@umk.pl}\\
~\\
\end{center}

\begin{abstract}  

We report photometry results of a frequently outbursting dwarf nova, ER Ursae Majoris. To measure the outburst parameters of the system, we carried out analyses of the light curve, periodograms, and $ O-C$ diagrams. We investigated the system's behaviour using the ground-based optical data and the Transiting Exoplanet Survey Satellite data. During these observation runs, we scrutinised three superoutbursts and several normal outbursts. We detected ordinary and late superhumps during each of the investigated superoutbursts. We derived the period excess value $\epsilon \approx 3.0(1)\% $. This suggests that over the last 30 years, ER UMa has not shifted on the evolutionary path toward period-bounce objects. Between 1992 and 2022, the interval between two successive superoutbursts (the supercycle length) changed significantly from 42.1 days to 59.6 days, which indicates that the mean mass-transfer rate of ER UMa has been decreasing over this period.

\noindent {\bf Key words:} \textit{Stars: individual:  ER UMa - binaries: 
close - novae, cataclysmic variables}

\end{abstract}


\section{Introduction}

Cataclysmic variables (CVs) are close binary systems with a white dwarf (the primary) and usually a low-mass main-sequence donor (the secondary), which fills its Roche lobe and loses its mass through the inner Lagrangian point. The transferred matter forms an accretion disk around the primary (Warner, 1995).

Among CVs, their subclass of dwarf novae (DNe) presents frequent rebrightenings with large amplitudes, and this allows us to study accretion disk properties under different physical conditions. The SU UMa-type DNe exhibit two main patterns of rebrightenings, denoted as superoutbursts and outbursts. Superoutbursts are about one magnitude brighter and about ten times less frequent than regular outbursts. The supercycle length, defined as the time between two successive superoutbursts, is specific to each system. In the light curves of SU UMa-type stars, photometric tooth-shaped modulations known as superhumps are visible. Stars with positive superhumps exhibit oscillations with periods a few per cent longer than their orbital periods. For negative superhumps, one can detect periods slightly shorter than orbital periods, (for review: Hellier, 2001). 
 
ER Ursae Majoris (ER UMa) was discovered as an ultraviolet excess object by  Green, Schmidt and Liebert (1986), and classified as a member of the SU UMa group by Kato and Kunjaya (1995). The orbital period of this system was estimated by Thorstensen et al. (1997) as 0.06366(3) days. The following analysis was presented by Gao, Li, and Wu (1999). They suspected the presence of a negative superhump period of $0.0589(7)$ days and positive superhump periods of $0.0653(10) - 0.0642(4)$ days. Later, several publications focused on the superhump periods of ER UMa, which were presented by Ohshima et al. (2012, 2014) and Kato et al. (2013). Additionally, Otulakowska-Hypka and Olech (2013), and Zemko, Kato and Shugarov (2013) investigated the supercycle length of ER UMa, obtaining $\dot{P}_{sc\_OHO}=12.7(1.9)\times10^{-4}$ and $\dot{P}_{sc\_ZKS}=6.7(6)\times10^{-4}$, respectively. However, Bean (2021) has recently pointed out significant discrepancies in the values of supercycle lengths, which is the motivation for this work. 


\section{Observations and data reduction}

We conducted observations of ER UMa between March 1, 2022, and May 23, 2022. The observations covered 26 nights, yielding 4 344 useful measurements. The exposure times ranged from 30 to 90 seconds, depending on the weather and the actual brightness of the star. In total, we gathered 59.55 hours of observations. The typical accuracy of our measurements varied between 0.002 mag and 0.110 mag. The median value of the photometric errors was 0.006 mag. Table 1 presents our ER UMa observations journal. 

All the data were collected at the Institute of Astronomy of the Nicolaus Copernicus University (the Piwnice Observatory near Toru\'{n}). ER UMa was observed using the 60-cm Cassegrain telescope equipped with the FLI ML16803 CCD camera, which provides an 18' x 18' field of view. The object was monitored in white light (a clear filter). 

The science frames were subjected to standard reduction and photometry procedures implemented in the AstroImageJ software (Collins et al., 2017). Each image was corrected for bias and thermal noise and normalised by its flat field. Light curves were obtained using the aperture photometry method with the aperture radius and ensemble of comparison stars optimised in trial iterations. The built-in procedure in AstroImageJ was used to convert timestamps into Barycentric Julian Ephemeris Dates (BJD$_\text{TDB}$). 

In Fig.\,1, the map of the observed region is presented, where ER UMa is marked as T1, and the three comparison stars are marked as C2, C3 and C4, respectively. The equatorial coordinates of the comparison star C4 (RA=$09^\text{h}47^\text{m}03^\text{s}.485$, Dec=$+51^\text{o}55'09''.338$) are taken from the Gaia Data Release 3 (Gaia DR3, Gaia Collaboration 2021). Gaia DR3 also provided the parallax of ER UMa measured as $2.8039\pm0.0205$ mas, corresponding to a distance of approximately 357 parsecs. The renormalised unit weight error (RUWE) of 1.10 suggests a reliable fit of the astrometric solution.

\begin{table*}[!ht]
\label{Table1}
 \centering
 \begin{small}
  \caption{Journal of our CCD observations of ER UMa.}
\bigskip
  \begin{tabular}{@{}|c|c|c|c|c|@{}}
  \hline
Date of   & Time of start   & Length     & Number of      \\
 2022     & BJD - 2450000  & of run [h] & frames               \\
\hline
March 01 	&	 9640.260295 	&	 1.1 	&	82	\\
March 02 	&	 9641.248073 	&	 2.0 	&	160	\\
March 10 	&	 9649.310984 	&	 0.5 	&	45	\\
March 11 	&	 9650.349745 	&	 0.7 	&	55	\\
March 12 	&	 9651.279485 	&	 2.0 	&	160	\\
March 13 	&	 9652.400515 	&	 1.9 	&	143	\\
March 20 	&	 9659.344502 	&	 0.5 	&	45	 \\
March 21 	&	 9660.308096 	&	 7.9 	&	751	\\
March 22 	&	 9661.290440 	&	 8.7 	&	493	 \\
March 23 	&	 9662.259688 	&	 9.2 	&	527	 \\
March 24 	&	 9663.268519 	&	 8.4 	&	571	\\
March 25 	&	 9664.281047 	&	 2.2 	&	165	\\
March 26 	&	 9665.313258 	&	 6.7 	&	509	\\
\hline								
April 11 	&	 9681.342899 	&	 1.0 	&	80	 \\
April 12 	&	 9682.439439 	&	 0.25 	&	20	\\
April 17 	&	 9687.373050 	&	 0.5 	&	37	\\
April 22 	&	 9692.330631 	&	 0.5 	&	40	\\
April 28 	&	 9698.349855 	&	 0.5 	&	37	\\
April 29 	&	 9699.482124 	&	 0.4 	&	31	\\
May 02 	&	 9702.406719 	&	 0.5 	&	38	 \\
May 06 	&	 9706.528177 	&	 0.5 	&	40	 \\
May 08 	&	 9708.379769 	&	 0.1 	&	9	\\
May 09 	&	 9709.408576 	&	 0.5 	&	30	\\
May 17 	&	 9717.368889 	&	 2.3 	&	135	 \\
May 18 	&	 9718.459919 	&	 0.1 	&	106	 \\
May 23 	&	 9723.455567 	&	 0.6 	&	35	\\ 
\hline
\end{tabular}
 \end{small}
\end{table*}

\begin{figure}[!ht]
\centering
\includegraphics[width=0.55\textwidth]{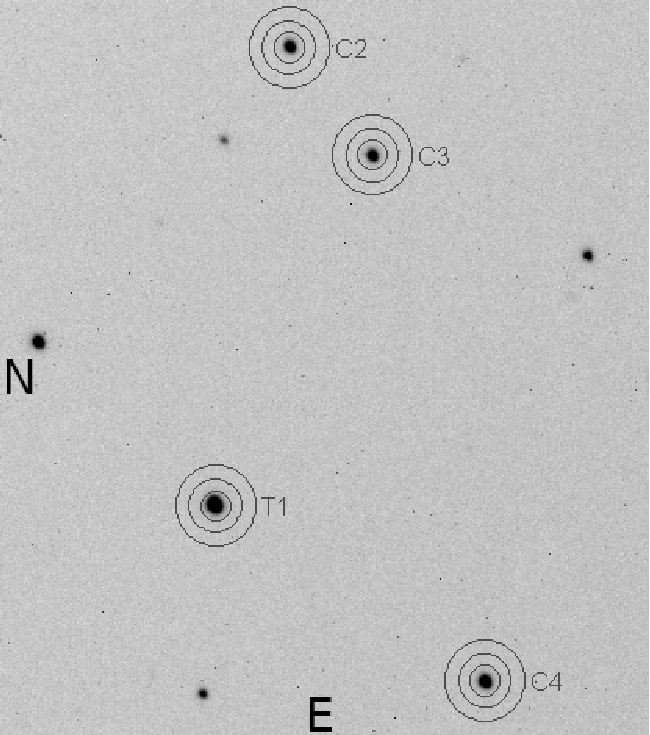}
    \caption {Finding chart of ER UMa, marked as T1. The positions of three comparison stars are also shown as C2, C3 and C4, respectively. The field of view is about ~5'$\times$~5'. North is left, and east is down.}
\end{figure}  


\section{Light curves}

\subsection{Global photometric behaviour}

For convenience, we will use only a day number, BJD-2450000 [d], to refer to our observations from now on. In our study, we supplemented our data from the Piwnice Observatory with publicly available observations and measurements from the TESS satellite (sectors 21, 48, 74 and 75). Below, one can find the description of each dataset. 


\subsubsection{Publicly available observations}

We investigated light curves from publicly available services, including AAVSO (www.aavso.org), ASAS-SN (www.astronomy.ohio-state.edu/asassn/), and ZTF (www.ztf.caltech.edu). The collected data of ER UMa cover thirty years of observations from May 6, 1992, to May 31, 2022. In total, we acquired 121 889 observation points gathered by amateur and professional astronomers. To combine our observations with the downloaded publicly available data, we converted timestamps from Julian Dates in public light curves into Barycentric Julian Ephemeris Dates (BJD$_\text{TDB}$) using the software provided by Eastman, Siverd, and Gaudi (2010). We used all acquired observations only to investigate the supercycle and normal cycle lengths (Sec.\,4.1). For further detailed analysis of superhumps development (Sec.\,3.3), we chose light curves of the best nightly covered superoutbursts (denoted as 2011 Feb, 2011 May, 2012 Feb and 2012 May) from the AAVSO database. These examples of excellent quality public data are presented in Fig.\,2 (top panels), and details about them are given in Table 2. 

\begin{figure}[!ht]
   \centering
    \includegraphics[width=0.95\textwidth]{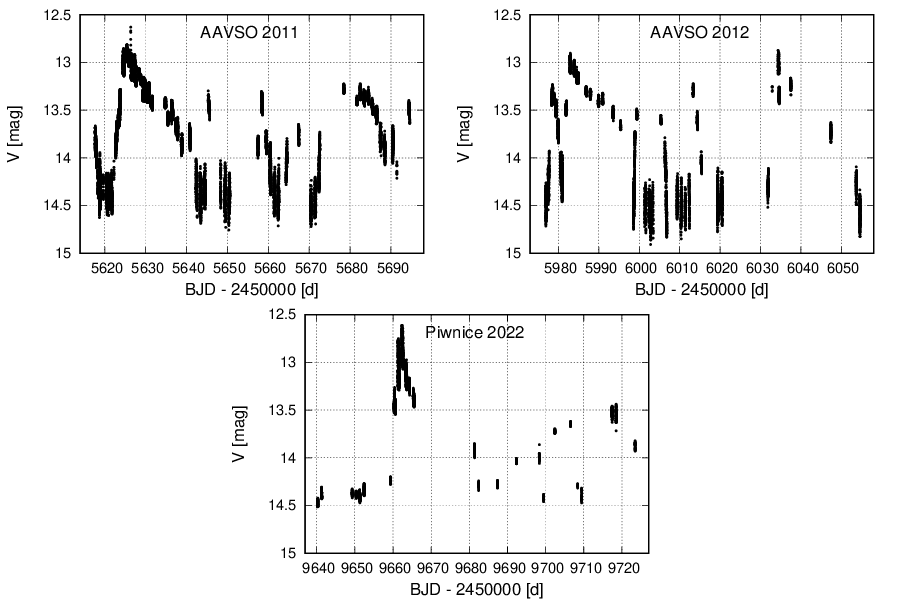}
      \caption {Global photometric behavior of ER UMa. \textit{Top panels}: Light curves from the AAVSO, \textit{bottom panel}: data from the Piwnice Observatory.  }
\end{figure} 


\subsubsection{Observations from the Piwnice Observatory}

The photometric behaviour of ER UMa, monitored in the Piwnice Observatory, is presented in the bottom panel of Fig.\,2. Relative magnitudes of the variable star were transformed to the $V$ filter magnitudes using the magnitudes of ER UMa estimated by amateur astronomers and published in AAVSO. Even though these transformations differ by no more than 0.1 mag, they are, in fact, rough estimates, which are used only to illustrate the general behaviour of the star. Details regarding two subsequent superoutbursts observed in the Piwnice Observatory are presented in Table 2. 


\subsubsection{TESS satellite data}

ER UMa was observed by the TESS satellite in four sectors: 21 (January 21, 2020, to February 18, 2020), 48 (January 28, 2022, to February 26, 2022), 74 (January 3, 2024, to January 30, 2024), and 75 (January 30, 2024, to February 26, 2024). During these runs, all exposures were acquired in the short cadence (SC) mode with exposure times of 2 min. The light curves were downloaded from the MAST Portal (https://mast.stsci.edu) and processed using standard procedures in the Lightkurve v1.9 package (Lightkurve Collaboration, 2018). We visually reviewed the data to eliminate measurements influenced by scattered light. In total, we acquired 57 086 exposures (1909.87 hours of observations). To investigate the general photometric behaviour of ER UMa, we converted the fluxes into $V$ filter magnitudes using Pogson's equation, along with an arbitrarily chosen constant shift, based on the AAVSO dataset as the template. Light curves of ER UMa from the TESS sectors are shown in four panels of Fig.\,3. Details about superoutbursts observed by the TESS are presented in Table 2.

\begin{figure}[!ht]
\centering
\includegraphics[width=0.95\textwidth]{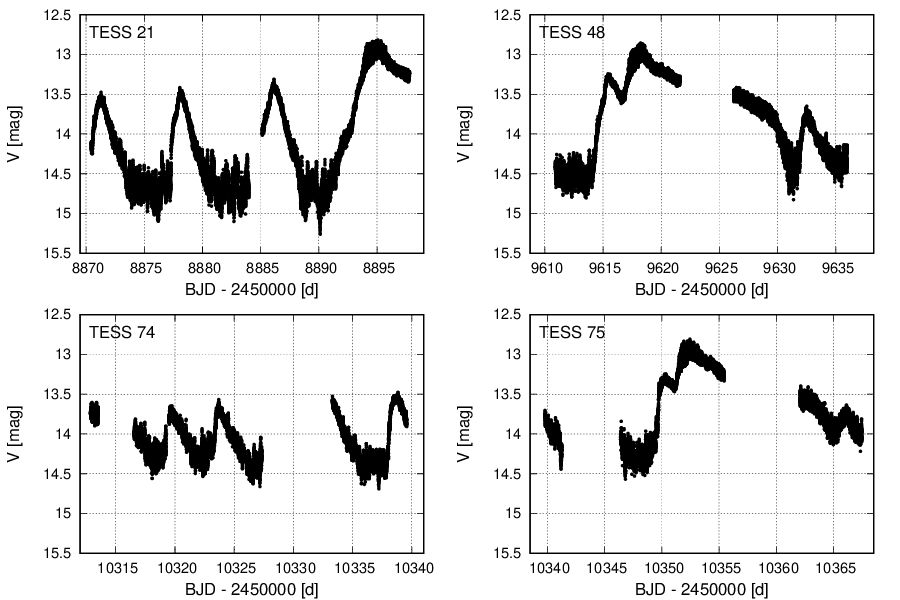}
\caption {Global photometric behavior of ER UMa, where data were provided by the TESS mission in four sectors: 21, 48, 74 and 75, shown on the subsequent panels.}
\end{figure}

\begin{table*}[!ht]
\label{Table2}
 \centering
 \begin{small}
  \caption{Properties of ER UMa superoutbursts.}
\bigskip
  \begin{tabular}{@{}|l|c|c|c|c|c|c|c|@{}}
  \hline
Date & Date of start   & Date of end   & Duration    & Max. brightness           & Amplitude &  Source  \\
      & BJD - 2450000 & BJD - 2450000      &      [d]        & in V [mag]    &  [mag]    & of data  \\
\hline
2011 Feb &	5621  & 5642    & 22    & 12.6  & 2.0 & AAVSO\\
2011 May &	5671  & 5692(1) & 22(1) & -     & -   & AAVSO\\
2012 Feb &	5981  & 5997(1) & 17(1) & 12.7  & 1.8 & AAVSO\\
2012 May &	6031  & 6050(2) & 20(2) & 12.7  & 1.7 & AAVSO\\
\hline
2022 Mar &	9659  & 9682    & 24	& 12.6  & 1.7 & Piwnice\\
2022 May &	9709  &  -  	&	-	&	-   & -   & Piwnice\\
\hline
2020 Feb &	8890  & -  	       &	& 12.7  & -   & TESS, (sec.\,21)\\
2022 Feb &  9615  & 9631	   & 17 & 12.8  & 2.0 & TESS (sec.\,48)\\
2024 Feb & 10349 & 10365     & 17 & 12.7  & 1.6 & TESS (sec.\,75)\\
\hline
\end{tabular}
 \end{small}
\end{table*}   

We compared the data presented in Table 2, and it is clear that between 2011 and 2024, the star reached maximum brightness values between $V \approx 12.6$ and $V \approx 12.8$ mag, and in quiescence, ER UMa faded to values between $V \approx 14.3$ and $V \approx 14.8$ mag. Hence, over the years, no significant changes in the superoutburst amplitude have been detected. However, the rate of change of the supercycle period of the variable changed, and we addressed this issue in Sec.\,4.1.


\subsection{Period analysis}

In our period search, we detrended each light curve of ER UMa by subtracting a straight line or parabola. This was done to remove the overall decreasing trend of superoutbursts and bring the mean value of all runs to zero. We decided to analyse each superoutburst, outburst or quiescence state of the object separately. We performed the period analysis with the Period04 code (Lenz and Breger, 2005) for each subset to calculate power spectra. Details are provided in the following subsections.

\subsubsection{Superoutbursts}

We performed the period analysis for the light curves of the five best-covered superoutbursts, denoted in Table 2 as 2011 Feb, 2022 Mar, 2020 Feb, 2022 Feb, and 2024 Feb. The resulting periodograms are shown in Fig.\,4. One prominent peak and its first harmonic corresponding to the superhump period are clearly visible for each superoutburst. For all five investigated superoutbursts, we found only a positive superhump period. Table 3 presents the detected periodicities in the light curves of ER UMa during its superoutbursts. Based on these periodogram statistics, the superhump period increased by $\sim 21$ seconds between 2011 and 2020. However, it also decreased by a value of $\sim 11$  seconds between 2020 and 2022. Therefore, these kinds of changes in the superhump period cannot be interpreted as any stable trend in the behaviour of ER UMa. 

\begin{table*}[ht]
\label{Table3}
 \centering
 \begin{small}
  \caption{Values of frequencies and periods determined for the five superoutbursts of ER UMa.}
\bigskip
  \begin{tabular}{@{}|l|l|l|c|@{}}
  \hline
Date of superoutburst        & Frequency   & Superhump        & Type of\\
(source of data)             & [c/d]       & period [d]       &  superhumps\\                
\hline 
2011 Feb (AAVSO)   & $f_{sh1} = 15.236(4) $ & $P_{sh1} = 0.06560(2)$ & positive \\               
2022 Mar (Piwnice) & $f_{sh2} = 15.200(20)$ & $P_{sh2} = 0.06579(9) $ & positive \\  
2020 Feb (TESS) & $f_{sh3} = 15.188(4)$ & $P_{sh3} = 0.06584(2) $ & positive\\  
2022 Feb (TESS) & $f_{sh4} = 15.218(3)$ & $P_{sh4} = 0.06571(1) $ & positive \\  
2024 Feb (TESS) & $f_{sh5} = 15.194(3) $ & $P_{sh5} = 0.06582(2) $ & positive  \\  
\hline
\end{tabular}
 \end{small}
\end{table*} 


In the light curve of ER UMa, there is a clear presence of tooth-shaped superhumps during superoutbursts, and Fig.\,5 shows the example of the evolution of the superhumps observed during the 2022 Feb superoutburst. Since superhump periods can change significantly during superoutbursts, we decided to use our best-covered data sets, the 2022 Feb and the 2024 Feb superoutbursts, and divided each set into one-day blocks to compute the power spectra for each night separately. For both superoutbursts, we started from the first night of the superoutburst of the object, and hence BJD 9615 and BJD 10349. The resulting periodogram statistics for the one-day blocks are presented in Fig.\,6. 

We also checked the stability of the phase of superhumps as the potential presence of a phase reversal was mentioned for active dwarf novae of ER UMa-type stars by, e.g. Kato, Nogami and Masuda (2003), Olech et al. (2004). For the light curves of the Feb 2022 and Feb 2024 superoutbursts, we phased the data with the corresponding superhump period for each night, separately. In Fig.\,7 and Fig.\,8, the resulting plots are shown. In Table 5, all detected frequencies and periods are presented. During the first nights (BJD 9617 - BJD 9618 and BJD 10351 - BJD 10352), the ordinary superhumps with maxima at phase $\sim 1$ are clearly visible with no manifestation of late superhumps. From BJD 9619 and BJD 10353, the situation changed completely, where one can see additional humps which begin to appear at the phase of $\sim 0.5$. 

During the first two nights covering outbursts - precursors preceding the 2022 Feb and the 2024 Feb superoutbursts (see first two left panels of Fig.\,6), no frequency was found on BJD 9615 - BJD 9616 and BJD 10349 - BJD 10350. In the following nights of maximum brightness of superoutbursts (BJD 9617 - BJD 9621 and BJD 10351 - BJD 10354), we found a strong peak at frequency $f$, but also at $2f$. This feature can be interpreted as the modulation has a double wave structure, which is consistent with a visual inspection of the light curve for this interval (see panels corresponding to these BJD  dates on Fig.\,7 and Fig.\,8). During the following nights of the plateau phase of scrutinised superoutbursts (BJD 9626 - BJD 9629 and BJD 10362 - BJD 10363), we detected only the main peak located at $f$ and no signal manifestation at $2f$. During the last two nights of investigated superoutbursts (BJD 9630 - BJD 9631 and BJD 10364 - BJD 10365), ER UMa briefly entered the quiescence stage. However, during this time, we found the strongest peak at frequency $f$ but almost no signal at $2f$. Also, the phase-folded data for minimum brightness of the object are the most scattered (see Fig.\,7  and Fig.\,8). During echo-outbursts (BJD 9632 - BJD 9633 and BJD 10366), the detected signal was weaker and once again visible only at $f$. Although the detected peaks from both superoutbursts overlap nicely (see Fig.\,6), it should be noted that Table 4 reveals discrepancies ranging from $\sim5$ seconds (BJD 9631 vs. BJD 10365) to $\sim50$ seconds (BJD 9630 vs. BJD 10364). Notably, these differences in frequencies are observed during quiescence between superoutbursts and echo-outbursts in both cases. 

\begin{figure}[!ht]
   \centering
    \includegraphics[width=0.95\textwidth]{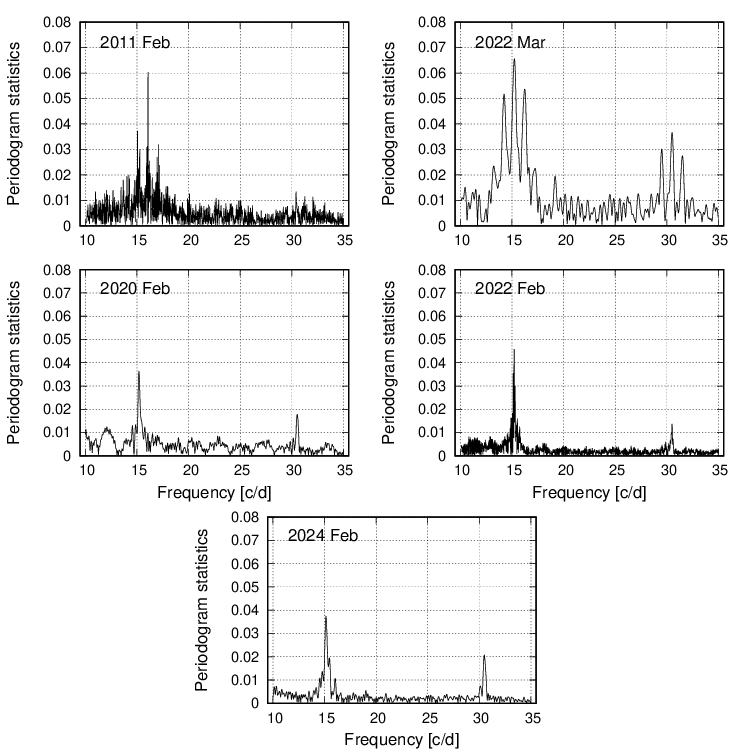}
      \caption {Periodograms of the ER UMa superoutbursts calculated based on the following datasets: \textit{top left:} AAVSO, \textit{top right:} Piwnice, \textit{middle left:} TESS, sec.\,21, \textit{middle right:} TESS, sec.\,48, and \textit{bottom:} TESS, sec.\,75.}
   \end{figure} 

\begin{figure}[!ht]
   \centering
    \includegraphics[width=0.95\textwidth]{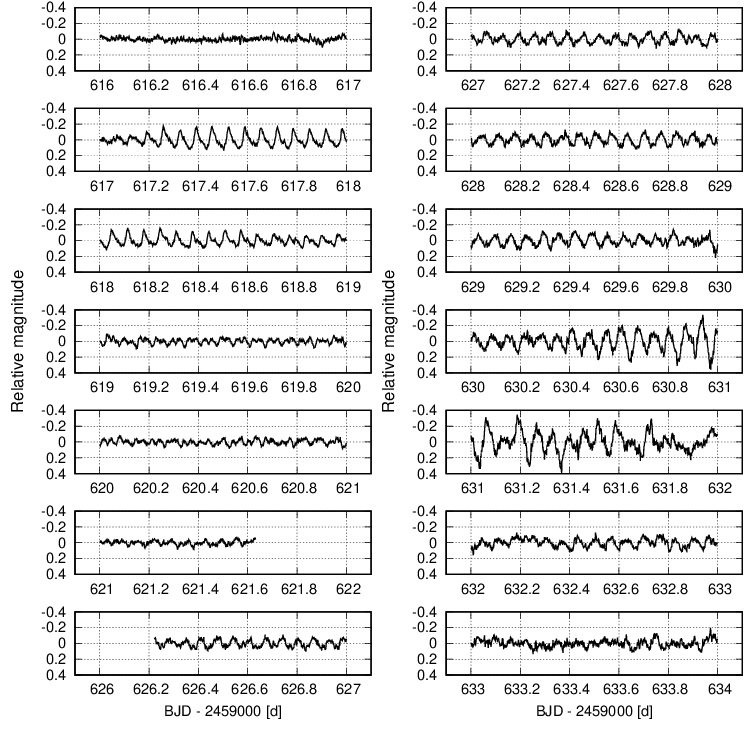}
      \caption {Superhumps observed during the 2022 Feb superoutburst of ER UMa (TESS, sec.\,48). }
   \end{figure}

\begin{figure}[!ht]
   \centering
    \includegraphics[width=0.95\textwidth]{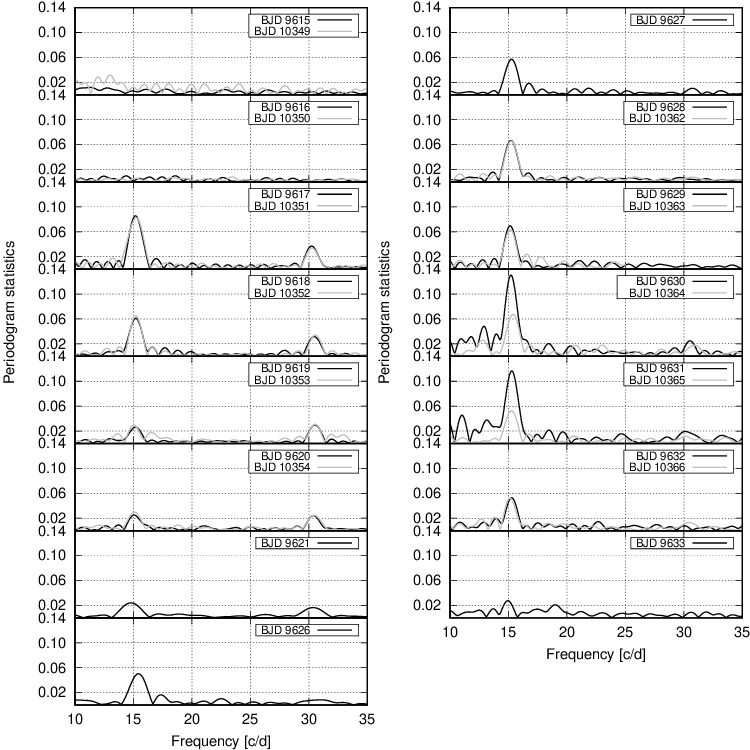}
      \caption {Periodogram statistics for the one-day blocks of the 2022 Feb (black line) and the 2024 Feb (grey line) superoutbursts of ER UMa (TESS, sec.\,48 and sec.\,75, respectively).}
   \end{figure} 

\begin{table*}[!ht]
\label{Table4}
 \centering
 \begin{small}
  \caption{Values of frequencies and periods determined for one-day blocks of the 2022 Feb and the 2024 Feb superoutbursts.}
\smallskip
  \begin{tabular}{@{}|c|c|c|c|c|c|@{}}
  \hline
   \multicolumn{3}{|c|}{The 2022 Feb superoutburst} & \multicolumn{3}{|c|}{The 2024 Feb superoutburst}\\
  \hline
Block           &	Frequency $f$ &	 Period  &  Block  & Frequency $f$ & Period\\
BJD - 2450000 	&	  [c/d]   &	  [d] &  BJD - 2450000 	& [c/d] & [d]  \\

\hline
9615 &     no detection       & -                   & 10349 & no detection & - \\
9616 &     no detection       &    -             & 10350 & no detection & - \\
9617	&	15.221(2)         & 0.06570(1) &  10351 & 15.257(2) & 0.06554(1) \\
9618	&	15.171(3)	      & 0.06592(1) &  10352 & 15.207(4) & 0.06576(2) \\
9619	&	15.171(5)	      & 0.06592(2) &  10353 & 15.057(6) & 0.06641(3) \\
9620	&	15.071(6)         & 0.06635(3) &  10354 & 15.107(6) & 0.06620(3) \\ 
9621	&	14.830(5)	      & 0.06743(2) &  10355 & no data & -  \\ 
9622-9625 &	no data	          &      -     &  10356-10359 & no data & - \\ 
9626 &	15.429(5)             & 0.06481(2) &  10360 & no data & -  \\
9627 &	   15.222(4)          & 0.06569(2) &  10361 & no data & - \\
9628 & 15.222(3) & 0.06569(1)              &  10362 & 15.258(3) & 0.06554(1) \\
9629 & 15.172(7) & 0.06591(3)              &  10363 & 15.157(5) & 0.06598(2) \\
9630 & 15.272(5) & 0.06548(2)              &  10364 & 15.408(6) & 0.06490(3) \\
9631 & 15.272(7) & 0.06548(3)              &  10365 & 15.258(6) & 0.06554(3) \\
9632 & 15.222(9)  & 0.06569(4)             &  10366 & 15.107(8) & 0.06620(4) \\
9633 & 14.922(9) & 0.06702(4)              &  10367 & no data & - \\
9634 & 15.172(8) & 0.06591(4)              &  10368 & no data & -\\  
\hline
\end{tabular}
 \end{small}
\label{tab0}
\end{table*}  

\begin{figure}[!ht]
   \centering
    \includegraphics[width=0.95\textwidth]{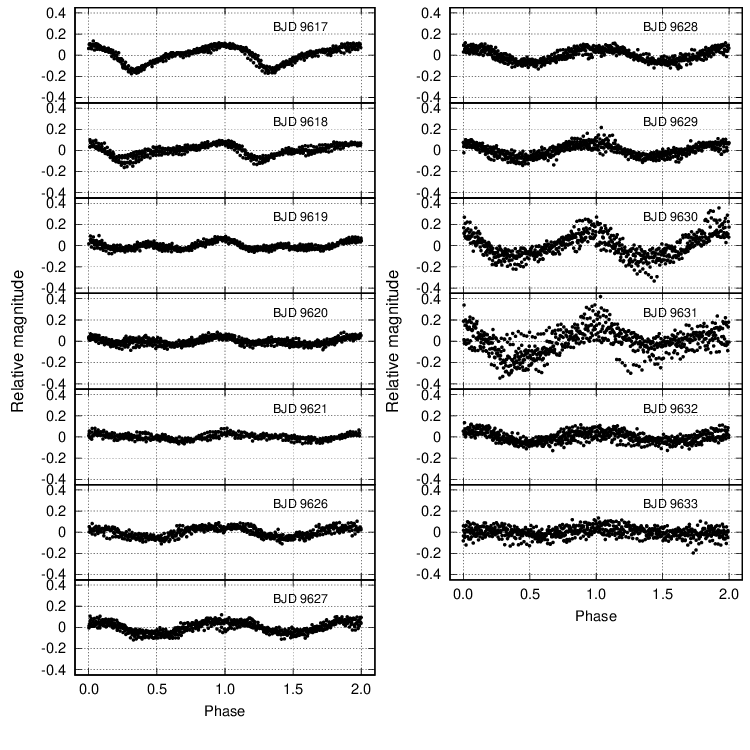}
      \caption {Evolution of superhumps during the 2022 Feb superoutburst. Panels display detrended light curves, phased with superhump periods given in Table 5, for each day separately.}
   \end{figure}

\begin{figure}[!ht]
   \centering
    \includegraphics[width=0.95\textwidth]{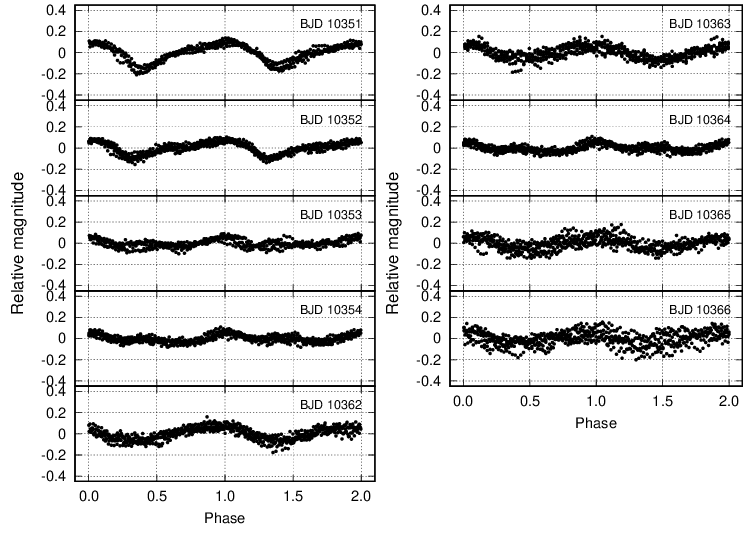}
      \caption {Evolution of superhumps during the 2024 Feb superoutburst. Panels display detrended light curves, phased with superhump periods given in Table 5, for each day separately.}
   \end{figure}


\subsubsection{Normal outbursts}

The TESS provided the best-quality data on ER UMa during its normal outbursts. Hence, we scrutinised three normal outbursts shown on the light curve displayed in Fig.\,3 (top, left panel), and one echo-outburst displayed in the light curve in Fig.\,3 (top, right panel). Because Zhao et al. (2006) and Zemko, Kato and Shugarov (2013) reported the occurrence of superhumps in normal outbursts of ER UMa, we looked for any signal corresponding to positive/negative superhump periods. We detected no manifestation of a negative superhump signal. The only peaks in periodograms, corresponding to the positive superhump period, we found on BJD 9632 and BJD 10366 during echo-outbursts after the Feb 2022 and the Feb 2024 superoutbursts (see  Fig.\,6, the right, second from bottom, panel).  

\subsubsection{Quiescence}

Once again, the TESS satellite provided the best uninterrupted light curves of ER UMa during its quiescence stage. Therefore, we investigated three data sets from sec.\,21 during the minimum brightness of the variable, right after three normal outbursts (Fig.\,3, top, left panel). We also analysed one set of pre-superoutburst data from sector 48 (Fig.\,3, top, right panel). In each case, there was no clear presence of any superhump frequency. The only manifestation of the positive superhump period was found on BJD 9631 and BJD 10365 (see Fig.\,6, right, third from bottom, panel). In both cases, ER UMa was transitioning from a superoutburst to an echo-outburst. 


\subsection{O-C Diagrams for superhumps}

An O-C diagram is one of the most popular methods for checking the stability of the superhump period and determining its value. For the 2011 Feb superoutburst, the $O-C$ was calculated by Ohshima et al. (2014), and they obtained $P_{sh6} = 0.65619$ days. In the case of the 2022 Mar superoutburst, the value of the superhump period $P_{sh7} = 0.06585(12)$ days was derived by Kirpluk et al. (2024). Therefore, below, we present the $O-C$ analysis dedicated to the 2020 Feb, the 2022 Feb, and the 2024 Feb superoutbursts monitored by TESS (sec.\,21, sec.\,48, and sec.\,75, respectively).


\subsubsection{The 2020 February Superoutburst}

In total, we identified 93 superhumps maxima in the light curves of ER UMa during its 2020 Feb superoutburst (TESS, sec.\,21). Worth noting is that, apart from the standard integer values of cycle numbers, we found 33 moments of maxima shifted by a half of a cycle with respect to these ordinary ones. The cycle number $E$, times and errors, and $O-C$ values, calculated for ordinary and late superhumps, are listed in Table 7. The corresponding $O-C$ diagram is shown in Fig.\,9, where black dots represent maxima of ordinary superhumps and open squares represent maxima of shifted superhumps. The example light curve of ordinary superhumps is shown in the upper window in Fig.\,9, and the simultaneous presence of both types of superhumps is visible in the lower window in Fig.\,9.

\begin{figure}[!ht]
   \centering
    \includegraphics[width=0.95\textwidth]{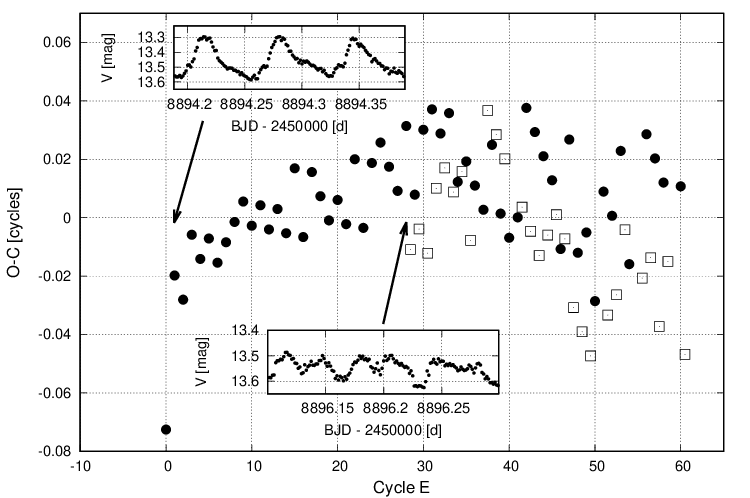}
      \caption {The $O-C$ for superhump maxima detected in TESS (sec.\,21) during the 2020 Feb superoutburst. Black dots represent typical superhumps, and the upper window displays an example of the light curve of these superhumps. Open squares represent late superhumps, and the lower window displays an example of a light curve for the simultaneous presence of typical and late superhumps.}
\end{figure} 

To check the stability of superhump periods, we performed an $O-C$ analysis for three sets of data: ordinary superhumps, late superhumps and both types of superhumps altogether. Least squares linear fits to these data gave the following ephemerides, respectively:  

\begin{equation}
\text{BJD}_{\text{max1}} = 8893.8211(55) + 0.06558(1) \times E,
\end{equation}

\begin{equation}
\text{BJD}_{\text{max1+0.5}} = 8893.8254(20) + 0.06545(4)\times E,
\end{equation}

\begin{equation}
\text{BJD}_{\text{max1}\_\text{all}} = 8893.8218(52) + 0.06555(1) \times E.
\end{equation}

Additionally, the second-order polynomial fit was calculated for the moments of maxima of all superhumps, and the corresponding ephemeris was obtained as:

\begin{equation}
\text{BJD}_{\text{max}} = 8893.8198(8) + 0.06572(6) \times E - 3.0(9) \times 10^{-6} \times E^2,
\end{equation}

\noindent
indicating that the period was decreasing with the rate of $\dot{P_1} = -9.1(2.7) \times 10^{-5}$. In Fig.\,11 (top left panel), the solid line corresponds to the ephemeris (4). All superhump periods derived from ephemerides (1), (2) and (3) are presented in Table 5.


\subsubsection{The 2022 February Superoutburst}

In the light curve of ER UMa during its 2022 Feb superoutburst (TESS, sec.\,48), we detected 197 superhump maxima. In particular, 68 moments of ordinary maxima, and 129 moments of maxima shifted by a half-cycle were identified. In Table 8, the cycle number $E$, times and errors, and $O-C$ values are presented, derived for ordinary and late superhumps. The resulting $O-C$ diagram is displayed in the top panel of Fig.\,10.

The linear fits to moments of maxima for ordinary superhumps, late superhumps and both types of superhumps altogether resulted in the following ephemerides, respectively:

\begin{equation}
\text{BJD}_{\text{max2}} = 9617.1296(5) + 0.06559(1) \times E,
\end{equation}

\begin{equation}
\text{BJD}_{\text{max2+0.5}} = 9617.1300(7) + 0.065586(5) \times E,
\end{equation}

\begin{equation}
\text{BJD}_{\text{max2}\_\text{all}} = 9617.1298(3) + 0.065587(4) \times E.
\end{equation}

This time, the second-order polynomial fit was calculated for both types of moments of maxima for cycles from $E=0$ to $E=59.5$, where a clear parabolic shape was visible. As a result, the corresponding ephemeris was obtained as:

\begin{equation}
\text{BJD}_{\text{max}} = 9617.1282(7) + 0.06571(5) \times E - 1.8(7) \times 10^{-6} \times E^2,
\end{equation}

\noindent
indicating that the period was decreasing with the rate of $\dot{P_2} = -5.5(2.1) \times 10^{-5}$. In the top, right panel of Fig.\,11, the solid line corresponds to the ephemeris (8). In Table 5, superhump periods are listed, obtained from ephemerides (5), (6), and (7).


\subsubsection{The 2024 February Superoutburst}

In the light curve of ER UMa during its 2024 Feb superoutburst, we determined 137 times of maxima (54 ordinary and 83 late superhumps), which are listed in Table 9 together with their cycle numbers, $E$, and $O-C$ values. The resulting $O-C$ diagram is presented in the bottom panel of Fig.\,10. The cycle numbers $E$ and times of maxima were fitted with the linear ephemerides for ordinary superhumps, late superhumps, and both types of superhumps altogether, respectively:

\begin{equation}
\text{BJD}_{\text{max3}} = 10351.5296(10) + 0.06549(4) \times E,
\end{equation}

\begin{equation}
\text{BJD}_{\text{max3+0.5}} = 10351.5245(13) + 0.065594(8) \times E,
\end{equation}

\begin{equation}
\text{BJD}_{\text{max3}\_\text{all}} = 10351.5272(6) + 0.065578(5) \times E.
\end{equation}

Once again, the second-order polynomial fit was calculated for both types of moments of maxima for cycles from $E=0$ to $E=68$, where a clear parabolic shape was visible. The corresponding ephemeris was obtained:

\begin{equation}
\text{BJD}_{\text{max}} = 10351.5276(14) + 0.06572(13) \times E - 4.1(2.1) \times 10^{-6} \times E^2,
\end{equation}

\noindent
indicating that the period was decreasing with the rate of $\dot{P_3} = -12.5(6.4) \times 10^{-5}$. In the bottom panel of Fig.\,11, the solid line corresponds to the ephemeris (12). In Table 5, superhump periods are listed, obtained from ephemerides (9), (10), and (11).

\begin{figure}[!ht]
   \centering
    \includegraphics[width=0.95\textwidth]{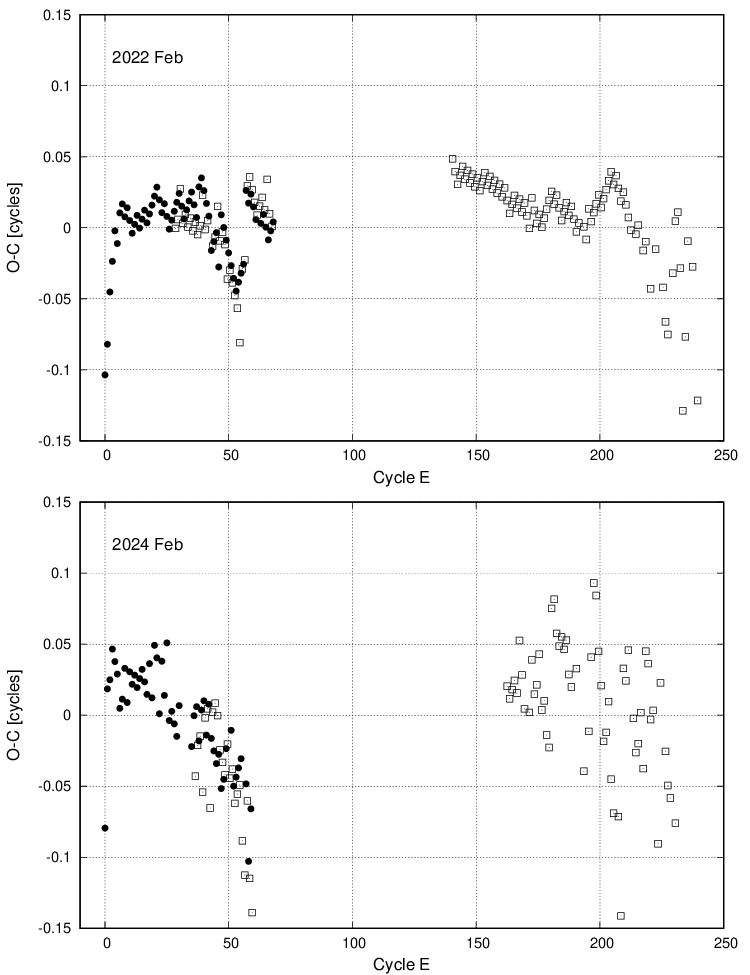}
      \caption {$O-C$ diagrams for superhump maxima detected in TESS (upper panel - sec.\,48, lower panel - sec.\,75), where black dots correspond to integer values of cycle numbers for ordinary superhumps, and the open squares present values shifted by a half of a cycle for late superhumps.}
\end{figure} 

\begin{figure}[!ht]
   \centering
    \includegraphics[width=0.95\textwidth]{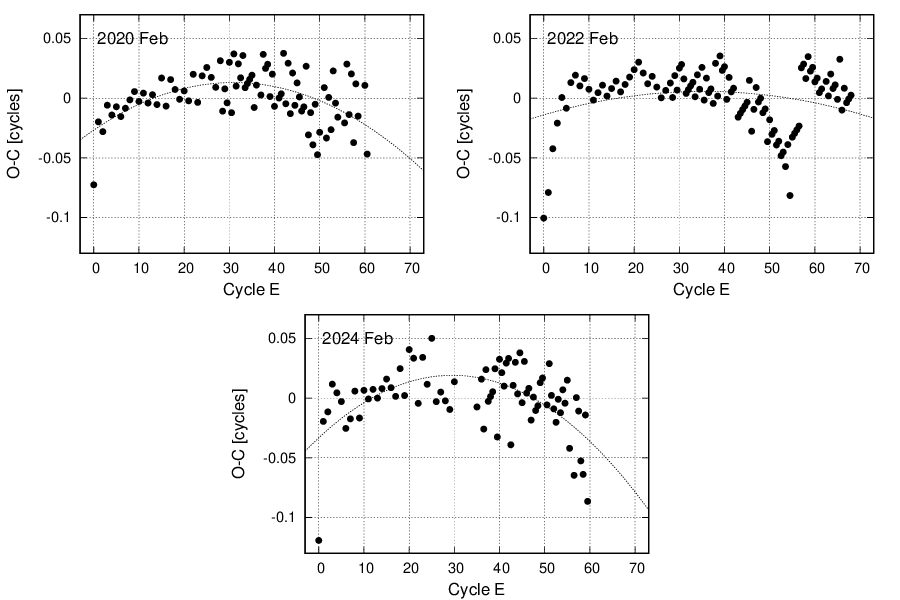}
      \caption {$O-C$ diagrams for the 2020 Feb (upper left panel), the 2022 Feb (upper right panel) and the 2024 Feb (lower panel) superoutbursts in ER UMa with the second-order polynomial fits (dotted black curves). }
\end{figure}

\begin{table*}[!ht]
\label{Table5}
\centering
 \begin{small}
  \caption{Values of periods determined for the 2020 Feb, the 2022 Feb and the 2024 Feb superoutbursts.}
\smallskip
  \begin{tabular}{@{}|l|c|c|l|@{}}
  \hline
    Superoutburst &	Period  &	 $E$  & Type of superhumps  \\
                  &	[d]     &	      &   \\   
\hline
2020 Feb & $P_{sh8} = 0.06558(1)$ & $0 \leqslant E \leqslant 60$ & ordinary \\
2020 Feb & $P_{sh9} = 0.06545(4)$ & $28.5 \leqslant E \leqslant 60.5$ & late\\
2020 Feb & $P_{sh10} = 0.06555(1)$ & $0 \leqslant E \leqslant 60.5$ & both\\
\hline
2022 Feb & $P_{sh11} = 0.06559(1)$ & $0 \leqslant E \leqslant 68$ & ordinary \\
2022 Feb & $P_{sh12} = 0.065586(5) $ & $28.5 \leqslant E \leqslant 67.5$, $140.5 \leqslant E \leqslant 239.5$ & late \\
2022 Feb & $P_{sh13} = 0.065587(4)$ & $0 \leqslant E \leqslant 239.5 $ & both \\
\hline
2024 Feb & $P_{sh14} = 0.06549(4)$ & $0 \leqslant E \leqslant 59$ & ordinary \\
2024 Feb & $P_{sh15} = 0.065594(8)$ & $36.5 \leqslant E \leqslant 59.5$, $162.5 \leqslant E \leqslant 230.5$ & late \\
2024 Feb & $P_{sh16} = 0.065578(5)$ & $0 \leqslant E \leqslant 230.5$& both \\
\hline
\end{tabular}
 \end{small}
\end{table*}  


\section{Discussion}

\subsection{Supercycle length}

The supercycle length is one of the most fundamental properties of SU UMa-type stars because the increase in the supercycle length corresponds to the decrease in accretion rate between components in CVs. As a result, this allows us to predict the future evolution of these variables. The first supercycle length measurement of ER UMa was done by Otulakowska-Hypka and Olech (2013). They used publicly available light curves from 1994 to 2010, averaging individual ER UMa supercycle periods into five bins of equal time range. They reported changes in superoutburst activity of ER UMa ranging from 42.7 days to 51.5 days, with the period derivative of $\dot{P}_{sc\_OHO} = 12.7\pm 1.9 \times 10^{-4}$. Almost immediately, Zemko, Kato and Shugarov (2013) published the second, more detailed analysis dedicated to the supercycle length of ER UMa. They utilised the same publicly available databases and data, covering nearly the same period from 1992 to 2012, as Otulakowska-Hypka and Olech (2013). However, no averaging was applied to the data measurements presented by Zemko, Kato and Shugarov (2013). What they noted was a change in the supercycle length ranging from 43.6 days to 59.2 days with the rate of supercycle growth of $\dot{P}_{sc\_ZKS}=6.7(6) \times 10^{-4}$. Hence, Zemko, Kato, and Shugarov (2013) estimated the rate of change of the supercycle period to be approximately half the value reported by Otulakowska-Hypka and Olech (2013). Recently, Bean (2021) presented an updated analysis of the $P_{sc}$ supplementing the results of Zemko, Kato and Shugarov (2013) with a new set of observations from 2012 to 2021. Therefore, the latest rate of change of the supercycle period is $\dot{P}_{sc\_B} = 3.3\times10^{-4}$, one-quarter of the first published value $\dot{P}_{sc\_OHO}$.

To obtain the length of the supercycle from all our acquired observations, we computed the power spectra using the Period04 software for each light curve of two subsequent superoutbursts. We found that the most prominent peak of these power spectra always corresponded to the period of the supercycle length. To investigate our findings more precisely, we derived an analytical fit to the first superoutburst in each light curve using the Bezier curve. Then, we copied the Bezier profile forward to the succeeding superoutburst, using the corresponding supercycle period value from Period04. The example of this procedure is displayed in Fig.\,12, where the grey line is the Bezier fit, which was calculated for the first (left) superoutburst and copied to the second (middle) and the third (right) superoutbursts. The locations of the Bezier profile correspond to values of supercycle length periods derived from power spectra analysis. For the three superoutbursts shown in Fig.\,12, the length of supercycle significantly increased from $P_{sc1}= 44.0(0.5)$ days to $P_{sc2} = 51.1(1.0)$ days. This discontinuous change in supercycle lengths is a hallmark of ER UMa, as reported by Zemko, Kato, and Shugarov (2013). Fig.\,13 (top panel) shows the updated graph of the supercycle period. It includes previously published measurements (see: Table 1 in Zemko, Kato and Shugarov, 2013) as well as our calculations based on the whole set of data described in Sec.\,3.1. Due to a significant discrepancy between $\dot{P}_{sc\_{OHO}}$ and $\dot{P}_{sc\_{ZKS}}$ pointed out by Bean (2021), we decided to estimate the rate of change of the supercycle in two different ways. Not only did we use the original data, but we also binned data in the same way as was described in Otulakowska-Hypka and Olech (2013). Hence, on the bottom panel of Fig.\,13, there are two linear fits; the first represents all unbinned measurements (black dotted line), and the second (grey line) represents the binned dataset. The rates of change of the supercycle period of ER UMa for these two fits are  $\dot{P}_{sc\_unbin} = 5.7(1.0)\times10^{-4}$ days and  $\dot{P}_{sc\_bin} = 4.8(2.9)\times10^{-4}$ days, respectively. Table 6 includes all the measurements mentioned above. We postulate that the binning procedure did not significantly alter the rate of change of the supercycle period. Our $\dot{P}_{sc\_{unbin}}$ and $\dot{P}_{sc\_{bin}}$ are in agreement with values postulated by Zemko, Kato and Shugarov (2013). Moreover, we agree with Bean (2021) that, although the actual rate of mass transfer change is increasing, it occurs at a much slower rate than was postulated by Otulakowska-Hypka and Olech (2013). 

\begin{figure}[!ht]
   \centering
    \includegraphics[width=0.95\textwidth]{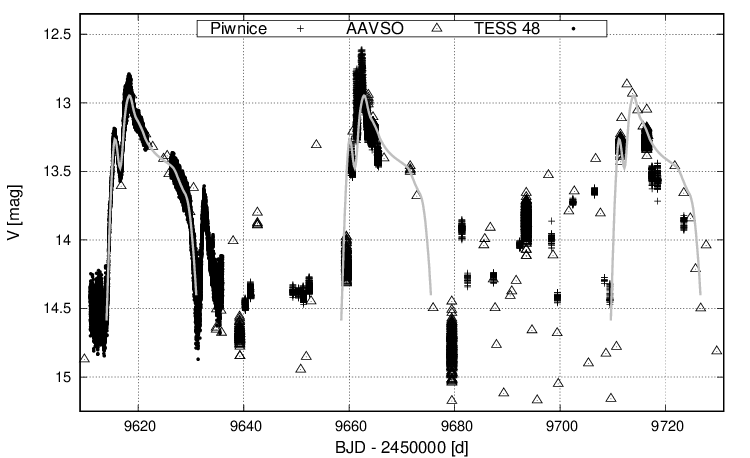}
      \caption {Observed superoutburst of ER UMa from the TESS sec.\,48 (left), succeeded by the superoutbursts (middle and right) from our observational campaign in the Piwnice Observatory and from the AAVSO archive. The supercycle periods estimated with Period04 were: for the left and middle superoutbursts, $P_{sc1} = 44.0(5)$ days, and for the middle and right superoutbursts, $P_{sc2} = 51.1(1.0)$ days. The grey lines indicate anticipated superoutbursts with the supercycle length $P_{sc1}$ and $P_{sc2}$, respectively.}
\end{figure}

\begin{figure}[!ht]
   \centering
    \includegraphics[width=0.95\textwidth]{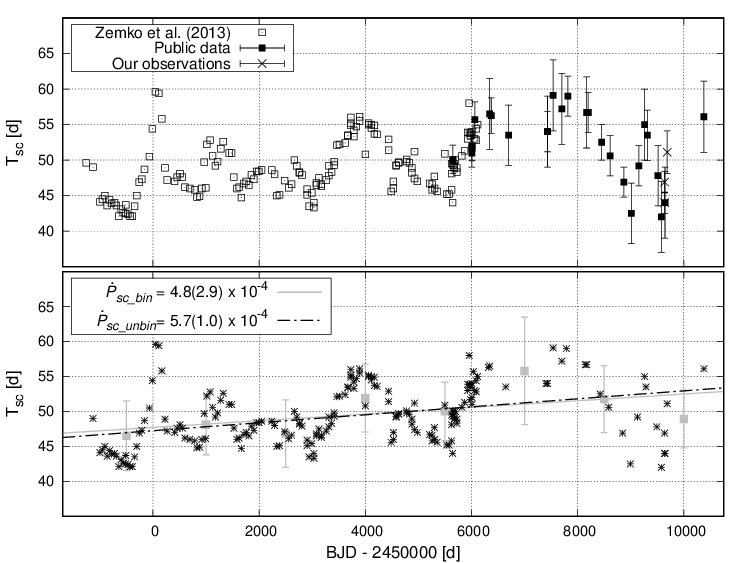}
      \caption {\textit{Top panel}: Variation of the supercycle lengths published by Zemko, Kato and Shugarov (2012) displayed as open squares, supplemented with publicly available data (black, filled squares), and updated by our measurements ("x" symbols). \textit{Bottom panel:} The same variation of the supercycle lengths plot with linear fits showing the rate of change of the supercycle period of ER UMa: the black dotted line corresponds to all unbinned measurements displayed as black stars, and the grey line corresponds to binned data marked as grey, filled squares.}
\end{figure}

We also investigated the normal cycle length. Kato and Kunjaya (1995) detected an extremely short recurrence time of 4 days for normal outbursts. Zemko, Kato, and Shugarov (2013) postulated that the normal cycle length, $T_c$, varies from 4 to 7 days. Once again, we verified our data set using the Period04 software, and we can confirm the previously published values of $T_c$. Also, no periodicity or clearly visible changes in the rate of normal cycle length were detected.

\begin{table*}[!ht]
\label{Table6}
 \centering
 \begin{small}
  \caption{The rate of change of the supercycle period of ER UMa, together with the respective ranges of the supercycle lengths and the time span of observations.}
\smallskip
  \begin{tabular}{@{}|c|c|c|c|c|c|@{}}
  \hline
$\dot{P}_{sc}$ & $P^{min}_{sc}$   & $P^{max}_{sc}$   & Observations    & Source  & Type\\
               &          [d]     &    [d]           &  [year]        &    of data    &  of data     \\
\hline
 $\dot{P}_{sc\_OHO} = 12.7(1.9)\times10^{-4}$    & 42.7    & 51.5  & 1994-2010 & Otulakowska-Hypka and & binned\\
                                                 &         &       &
           &     Olech (2013                    & \\                                                 
  $\dot{P}_{sc\_ZKS} = 6.7(6)\times10^{-4}$    & 43.6    & 59.2  & 1992-2012 & Zemko, Kato, and Shugarov (2013) & unbinned\\

  $\dot{P}_{sc\_B} = 3.3\times10^{-4}$    & 42    & 61  & 2012-2020 &  Bean (2021) & unbinned\\

 $\dot{P}_{sc\_unbin} = 5.7(1.0)\times10^{-4}$ & 42.1 & 59.6 & 1992-2022 & This work & unbinned\\

  $\dot{P}_{sc\_bin} = 4.8(2.9)\times10^{-4}$ & 42.1 & 59.6 & 1992-2022 & This work & binned\\
\hline
\end{tabular}
 \end{small}
\end{table*}   


\subsection{Superhumps}

Knowing the values of superhump and orbital periods, we can conclude the evolution of ER UMa. CVs showing superhumps are known to follow the Stolz-Schoembs relations between the period excess $\epsilon$ , defined as $P_{sh  }/P_{orb}-1$, and the orbital period of the binary (Stolz \& Schoembs, 1984). To determine $\epsilon$, we took the value of $P_{orb} = 0.06366(3)$ presented by Thorstensen et al. (1997). We also took into account all the superhump period values from the $O-C$ analysis (see Table 5). The period excess $\epsilon$ turned out to be $\sim 3.0(1)\%$ in almost all cases. Taking into account that Olech, Rutkowski \& Schwarzenberg-Czerny (2009) obtained $\epsilon =3.0(2)\%$ for superhumps observed in ER UMa in 1995 (Kato, Nogami \& Masuda, 2003), it means that the period excess value of this very active DN has not changed within the last 30 years. In Fig.\,14 we presented this dependence for different types of CVs with a position of ER UMa drawn as a black square (our measurements) and as a grey x symbol (data provided by Kato, Nogami and Masuda, 2003).  

Next, by employing the empirical formula given by Patterson (1998):

\begin{equation}
    \epsilon = \dfrac{0.23q}{1+0.27q},                                                                             
\end{equation}

\noindent
we derived the mass ratio for ER UMa as equal to $q \approx 0.14$. Ohshima et al. (2014) estimated value of $q = 0.100(15)$ based on the photometric observations of ER UMa between 2011 and 2012. Scrutinising the diagram of $q$ vs. $P_{orb}$ (Fig. 14, Ohshima et al., 2014), we conclude that our value of $q$ is located closer to the standard evolutionary tracks provided by Knigge et al. (2011) than the one provided by Ohshima et al. (2014). Also, our results indicate that ER UMa is on the standard evolutionary stage, similar to typical CVs. Despite its high mass transfer rate, there is no evidence that the object evolves differently from other SU UMa-type dwarf novae. 

\begin{figure}[!ht]
   \centering
    \includegraphics[width=0.9\textwidth]{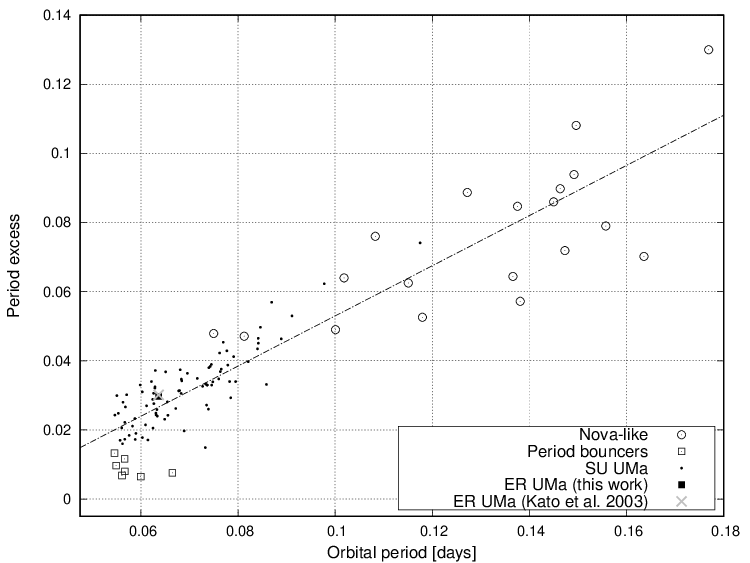}
      \caption {The dependence between the orbital period and the period excess for different types of cataclysmic stars. The old value for ER UMa is marked as a grey x symbol. New data for ER UMa is marked as a black square. All SU UMa stars are plotted with a black dots. Large open squares represent period bouncers, and circles correspond to nova-like variables.}
\end{figure}

Worth noting is that systems with extremely low mass ratio display early superhumps, i.e. RZ Leo with $q \sim 0.14$ (Ishioka et al., 2001). One can recognise early superhumps based on their period, which is always very close to the orbital period of the investigated variable. Additionally, they manifest as double-wave modulations, producing a power spectrum with stronger peaks at the second harmonics. Despite our detailed analysis of one-day blocks of superoutbursts data (Fig.\,6), we did not find any presence of early superhump period.

Another interesting feature was pointed by Smak (2016), who postulated the presence of secondary humps and the spurious phase jumps in the $O-C$ diagrams in a few active dwarf novae, and based on the data provided by Kato, Nogami and Masuda (2003, Fig. 1). Smak (20016) noticed the presence of the secondary humps from the first night of the superoutburst, and a jump of the phase of the primary maximum by $\sim 0.5$ during the 1995 superoutburst of ER UMa. Smak (2013) explained superhumps manifestation as being due to periodically modulated dissipation of the kinetic energy of the stream, and he discussed possible behavior of phases of maxima seen in the diagrams as the evolution of the disk outer structure. Although we detected a phase jump around $E \approx 57$ in the $O-C$ diagram for superhumps maxima during the 2022 Feb superoutburst (Fig.\,11 top right panel), there was no clear presence of a significant jump in the phase of the primary maximum in any other $O-C$ diagram (Fig.\,11 top left and bottom panels). In case of the occurrence of late maxima, shifted by a half of a cycle with respect to primary superhumps, is most probably a hallmark of dwarf novae with shorter orbital periods (Olech et al. 2004, Rutkowski et al. 2007). To explain the intricate shape of superoutburst oscillations, Smak (2004a,b, 2009) postulated the irradiation-modulated mass transfer model, which is especially valuable when the effects of the irradiation-enhanced mass outflow are the most significant in the CVs with the shortest orbital periods, and it could improve the modelling of the observed superhump amplitudes. However, the behaviour of ER UMa is very complex, and for the moment, it complicates our understanding of superhumps.

\bigskip

\noindent {\bf Acknowledgments.} ~It is our pleasant duty to acknowledge with thanks the use of the observations of ER UMa from the AAVSO, ASAS-SN and ZTF databases. Some of the observations have been obtained with the 60cm Cassegrain Telescope (TC60) in Piwnice Observatory, Institute of Astronomy of the Nicolaus Copernicus University in Toruń. 
(Poland).\\

\newpage

\begin{table*}[!ht]
\renewcommand\thetable{7. Part I}
\centering
 \begin{small}
  \caption{Times of maxima for ordinary superhumps in the light curve of ER UMa in the 2020 February superoutburst.}
\smallskip
  \begin{tabular}{@{}|l|c|l|r|l|c|l|r|@{}}
  \hline
$E$       &	Times of maxima  &	 Error  &  $O-C$ &  $E$         &	Times of maxima  &	 Error  &  $O-C$ \\
         &BJD$_{\text{max}} - 2450000$    &	  &  [cycles] &   &	BJD$_{\text{max}}-2450000$   &	   &  [cycles] \\

\hline
0 & 893.817 & 0.002 & -0.073 & 30 & 895.790 & 0.002 & 0.030 \\
1 & 893.886 & 0.002 & -0.020 & 31 & 895.856 & 0.002 & 0.037 \\
2 & 893.951 & 0.002 & -0.028 & 32 & 895.921 & 0.004 & 0.029 \\
3 & 894.018 & 0.004 & -0.006 & 33 & 895.987 & 0.002 & 0.036 \\
4 & 894.083 & 0.002 & -0.014 & 34 & 896.051 & 0.004 & 0.012 \\
5 & 894.149 & 0.002 & -0.007 & 35 & 896.117 & 0.002 & 0.019 \\
6 & 894.214 & 0.002 & -0.015 & 36 & 896.182 & 0.002 & 0.011 \\
7 & 894.280 & 0.002 & -0.008 & 37 & 896.247 & 0.002 & 0.003 \\
8 & 894.346 & 0.002 & -0.001 & 38 & 896.314 & 0.004 & 0.025 \\
9 & 894.412 & 0.002 & 0.006 & 39 & 896.378 & 0.002 & 0.001 \\
10 & 894.477 & 0.002 & -0.003 & 40 & 896.443 & 0.002 & -0.007 \\
11 & 894.543 & 0.004 & 0.004 & 41 & 896.509 & 0.004 & 0.000 \\
12 & 894.608 & 0.002 & -0.004 & 42 & 896.577 & 0.002 & 0.038 \\
13 & 894.674 & 0.004 & 0.003 & 43 & 896.642 & 0.002 & 0.029 \\
14 & 894.739 & 0.002 & -0.005 & 44 & 896.707 & 0.002 & 0.021 \\
15 & 894.806 & 0.002 & 0.017 & 45 & 896.772 & 0.004 & 0.013 \\
16 & 894.870 & 0.002 & -0.007 & 46 & 896.836 & 0.002 & -0.011 \\
17 & 894.937 & 0.002 & 0.016 & 47 & 896.904 & 0.002 & 0.027 \\
18 & 895.002 & 0.002 & 0.007 & 48 & 896.967 & 0.002 & -0.012 \\
19 & 895.067 & 0.002 & -0.001 & 49 & 897.033 & 0.002 & -0.005 \\
20 & 895.133 & 0.002 & 0.006 & 50 & 897.097 & 0.002 & -0.029 \\
21 & 895.198 & 0.002 & -0.002 & 51 & 897.165 & 0.002 & 0.009 \\
22 & 895.265 & 0.004 & 0.020 & 52 & 897.230 & 0.004 & 0.001 \\
23 & 895.329 & 0.002 & -0.004 & 53 & 897.297 & 0.002 & 0.023 \\
24 & 895.396 & 0.004 & 0.019 & 54 & 897.360 & 0.004 & -0.016 \\
25 & 895.462 & 0.002 & 0.026 & 56 & 897.494 & 0.002 & 0.029 \\
26 & 895.527 & 0.002 & 0.017 & 57 & 897.559 & 0.004 & 0.020 \\
27 & 895.592 & 0.002 & 0.009 & 58 & 897.624 & 0.004 & 0.012 \\
28 & 895.659 & 0.002 & 0.031 & 60 & 897.755 & 0.002 & 0.011 \\
29 & 895.723 & 0.002 & 0.008 &&&&\\      
\hline
\end{tabular}
 \end{small}
\end{table*}  

\begin{table*}[!ht]
\renewcommand\thetable{7. Part II}
 \centering
 \begin{small}
  \caption{Times of maxima for late superhumps in the light curve of ER UMa in the 2020 February superoutburst.}
\smallskip
  \begin{tabular}{@{}|l|c|l|r|l|c|l|r|@{}}
  \hline
$E$       &	Times of maxima  &	 Error  &  $O-C$ &  $E$         &	Times of maxima  &	 Error  &  $O-C$ \\
         &BJD$_{\text{max}} - 2450000$    &	  &  [cycles] &   &	BJD$_{\text{max}}-2450000$   &	   &  [cycles] \\

\hline
28.5 & 895.689 & 0.002 & -0.011 & 44.5 & 896.738 & 0.002 & -0.006 \\
29.5 & 895.755 & 0.002 & -0.004 & 45.5 & 896.804 & 0.002 & 0.001 \\
30.5 & 895.820 & 0.004 & -0.012 & 46.5 & 896.869 & 0.004 & -0.007 \\
31.5 & 895.887 & 0.002 & 0.010 & 47.5 & 896.933 & 0.002 & -0.031 \\
32.5 & 895.953 & 0.004 & 0.017 & 48.5 & 896.998 & 0.002 & -0.039 \\
33.5 & 896.018 & 0.004 & 0.009 & 49.5 & 897.063 & 0.002 & -0.047 \\
34.5 & 896.084 & 0.004 & 0.016 & 51.5 & 897.195 & 0.002 & -0.033 \\
35.5 & 896.148 & 0.002 & -0.008 & 52.5 & 897.261 & 0.002 & -0.026 \\
37.5 & 896.282 & 0.002 & 0.037 & 53.5 & 897.328 & 0.002 & -0.004 \\
38.5 & 896.347 & 0.002 & 0.028 & 55.5 & 897.458 & 0.002 & -0.021 \\
39.5 & 896.412 & 0.002 & 0.020 & 56.5 & 897.524 & 0.002 & -0.014 \\
41.5 & 896.542 & 0.002 & 0.004 & 57.5 & 897.588 & 0.002 & -0.037 \\
42.5 & 896.607 & 0.002 & -0.005 & 58.5 & 897.655 & 0.002 & -0.015 \\
43.5 & 896.672 & 0.004 & -0.013 & 60.5 & 897.784 & 0.002 & -0.047 \\
\hline
\end{tabular}
 \end{small}
 \end{table*}


\begin{table*}[!ht]
\renewcommand\thetable{8. Part I}
 \centering
 \begin{small}

  \caption{Times of maxima for ordinary superhumps in the light curve of ER UMa in the 2022 February superoutburst.}
\smallskip
  \begin{tabular}{@{}|l|c|l|r|l|c|l|r|@{}}
  \hline
$E$       & Times of maxima  &  Error  &  $O-C$ &  $E$         & Times of maxima  &  Error  &  $O-C$ \\
         &BJD$_{\text{max}} - 2450000$    &   &  [cycles] &   & BJD$_{\text{max}}-2450000$   &    &  [cycles] \\

\hline
0 & 9617.123 & 0.002 & -0.104 & 34 & 9619.361 & 0.002 & 0.019 \\
1 & 9617.190 & 0.002 & -0.082 & 35 & 9619.427 & 0.002 & 0.025 \\
2 & 9617.258 & 0.002 & -0.045 & 36 & 9619.492 & 0.002 & 0.016 \\
3 & 9617.325 & 0.002 & -0.024 & 37 & 9619.557 & 0.004 & 0.007 \\
4 & 9617.392 & 0.002 & -0.002 & 38 & 9619.624 & 0.002 & 0.029 \\
5 & 9617.457 & 0.002 & -0.011 & 39 & 9619.690 & 0.002 & 0.035 \\
6 & 9617.524 & 0.002 & 0.010 & 40 & 9619.755 & 0.002 & 0.026 \\
7 & 9617.590 & 0.002 & 0.017 & 41 & 9619.820 & 0.004 & 0.017 \\
8 & 9617.655 & 0.002 & 0.008 & 42 & 9619.885 & 0.004 & 0.008 \\
9 & 9617.721 & 0.002 & 0.014 & 43 & 9619.949 & 0.002 & -0.016 \\
10 & 9617.786 & 0.002 & 0.005 & 44 & 9620.015 & 0.002 & -0.010 \\
11 & 9617.851 & 0.002 & -0.004 & 45 & 9620.081 & 0.002 & -0.003 \\
12 & 9617.917 & 0.002 & 0.002 & 46 & 9620.145 & 0.002 & -0.028 \\
13 & 9617.983 & 0.002 & 0.009 & 47 & 9620.213 & 0.002 & 0.009 \\
14 & 9618.048 & 0.002 & -0.000 & 48 & 9620.278 & 0.002 & 0.000 \\
15 & 9618.114 & 0.002 & 0.006 & 49 & 9620.343 & 0.002 & -0.009 \\
16 & 9618.180 & 0.002 & 0.012 & 50 & 9620.408 & 0.002 & -0.018 \\
17 & 9618.245 & 0.002 & 0.003 & 51 & 9620.473 & 0.002 & -0.027 \\
18 & 9618.311 & 0.002 & 0.010 & 52 & 9620.538 & 0.002 & -0.036 \\
19 & 9618.377 & 0.002 & 0.016 & 53 & 9620.603 & 0.002 & -0.045 \\
20 & 9618.443 & 0.002 & 0.022 & 54 & 9620.669 & 0.002 & -0.038 \\
21 & 9618.509 & 0.002 & 0.029 & 55 & 9620.735 & 0.002 & -0.032 \\
22 & 9618.574 & 0.002 & 0.020 & 56 & 9620.801 & 0.005 & -0.026 \\
23 & 9618.639 & 0.002 & 0.011 & 57 & 9620.870 & 0.002 & 0.026 \\
24 & 9618.705 & 0.002 & 0.017 & 58 & 9620.935 & 0.002 & 0.017 \\
25 & 9618.770 & 0.002 & 0.008 & 59 & 9621.001 & 0.004 & 0.024 \\
26 & 9618.835 & 0.002 & -0.001 & 60 & 9621.066 & 0.004 & 0.015 \\
27 & 9618.901 & 0.002 & 0.005 & 61 & 9621.131 & 0.002 & 0.006 \\
28 & 9618.967 & 0.004 & 0.012 & 63 & 9621.262 & 0.002 & 0.003 \\
29 & 9619.033 & 0.004 & 0.018 & 64 & 9621.328 & 0.002 & 0.009 \\
30 & 9619.099 & 0.002 & 0.024 & 65 & 9621.393 & 0.002 & 0.000 \\
31 & 9619.164 & 0.002 & 0.015 & 66 & 9621.458 & 0.002 & -0.009 \\
32 & 9619.229 & 0.002 & 0.006 & 67 & 9621.524 & 0.002 & -0.002 \\
33 & 9619.295 & 0.002 & 0.013 & 68 & 9621.590 & 0.004 & 0.004 \\
\hline
\end{tabular}
     
 \end{small}
\end{table*}  


\begin{table*}[!ht]
\renewcommand\thetable{8. Part II}
 \centering
 \begin{small}

  \caption{Times of maxima for late superhumps in the light curve of ER UMa in the 2022 February superoutburst.}
\smallskip
  \begin{tabular}{@{}|l|c|l|r|l|c|l|r|@{}}
  \hline
$E$       & Times of maxima  &  Error  &  $O-C$ &  $E$         & Times of maxima  &  Error  &  $O-C$ \\
         &BJD$_{\text{max}} - 2450000$    &   &  [cycles] &   & BJD$_{\text{max}}-2450000$   &    &  [cycles] \\

\hline             
28.5 & 9618.999 & 0.002 & 0.000 & 49.5 & 9620.374 & 0.006 & -0.036 \\
29.5 & 9619.065 & 0.004 & 0.006 & 50.5 & 9620.440 & 0.004 & -0.030 \\
30.5 & 9619.132 & 0.004 & 0.027 & 51.5 & 9620.505 & 0.006 & -0.039 \\
31.5 & 9619.196 & 0.004 & 0.003 & 52.5 & 9620.570 & 0.004 & -0.048 \\
32.5 & 9619.262 & 0.004 & 0.009 & 53.5 & 9620.635 & 0.004 & -0.057 \\
33.5 & 9619.327 & 0.004 & 0.000 & 54.5 & 9620.699 & 0.004 & -0.081 \\
34.5 & 9619.393 & 0.002 & 0.007 & 55.5 & 9620.768 & 0.002 & -0.029 \\
35.5 & 9619.458 & 0.002 & -0.002 & 56.5 & 9620.834 & 0.006 & -0.023 \\
36.5 & 9619.524 & 0.002 & 0.004 & 57.5 & 9620.903 & 0.008 & 0.029 \\
37.5 & 9619.589 & 0.002 & -0.005 & 58.5 & 9620.969 & 0.006 & 0.036 \\
38.5 & 9619.655 & 0.004 & 0.001 & 59.5 & 9621.034 & 0.004 & 0.027 \\
39.5 & 9619.722 & 0.004 & 0.023 & 60.5 & 9621.099 & 0.004 & 0.018 \\
40.5 & 9619.786 & 0.004 & -0.001 & 61.5 & 9621.164 & 0.004 & 0.009 \\
41.5 & 9619.852 & 0.004 & 0.005 & 62.5 & 9621.230 & 0.004 & 0.015 \\
43.5 & 9619.982 & 0.004 & -0.013 & 63.5 & 9621.296 & 0.004 & 0.021 \\
44.5 & 9620.048 & 0.004 & -0.007 & 64.5 & 9621.361 & 0.006 & 0.013 \\
45.5 & 9620.115 & 0.004 & 0.015 & 65.5 & 9621.428 & 0.004 & 0.034 \\
46.5 & 9620.179 & 0.006 & -0.009 & 66.5 & 9621.492 & 0.006 & 0.010 \\
47.5 & 9620.245 & 0.004 & -0.003 & 67.5 & 9621.557 & 0.006 & 0.001 \\
48.5 & 9620.310 & 0.006 & -0.012 &  &  &  &  \\     
\hline
\end{tabular}
 \end{small}
\end{table*}  


\begin{table*}[!ht]
\renewcommand\thetable{8. Part III}
 \centering
 \begin{small}

  \caption{Times of maxima for late superhumps in the light curve of ER UMa in the 2022 February superoutburst.}
\smallskip
  \begin{tabular}{@{}|l|c|l|r|l|c|l|r|@{}}
  \hline
$E$       & Times of maxima  &  Error  &  $O-C$ &  $E$         & Times of maxima  &  Error  &  $O-C$ \\
         &BJD$_{\text{max}} - 2450000$    &   &  [cycles] &   & BJD$_{\text{max}}-2450000$   &    &  [cycles] \\
    
\hline
140.5 & 9626.348 & 0.006 & 0.048 & 185.5 & 9629.297 & 0.006 & 0.011 \\
141.5 & 9626.413 & 0.006 & 0.040 & 186.5 & 9629.363 & 0.004 & 0.018 \\
142.5 & 9626.478 & 0.006 & 0.031 & 187.5 & 9629.428 & 0.006 & 0.009 \\
143.5 & 9626.544 & 0.004 & 0.037 & 188.5 & 9629.494 & 0.006 & 0.015 \\
144.5 & 9626.610 & 0.006 & 0.043 & 189.5 & 9629.559 & 0.006 & 0.006 \\
145.5 & 9626.675 & 0.004 & 0.034 & 190.5 & 9629.624 & 0.004 & -0.003 \\
146.5 & 9626.741 & 0.004 & 0.040 & 191.5 & 9629.690 & 0.006 & 0.003 \\
147.5 & 9626.806 & 0.006 & 0.032 & 193.5 & 9629.821 & 0.004 & 0.001 \\
148.5 & 9626.872 & 0.006 & 0.038 & 194.5 & 9629.886 & 0.008 & -0.008 \\
149.5 & 9626.937 & 0.008 & 0.029 & 195.5 & 9629.953 & 0.008 & 0.013 \\
150.5 & 9627.003 & 0.008 & 0.035 & 196.5 & 9630.018 & 0.006 & 0.004 \\
151.5 & 9627.068 & 0.006 & 0.026 & 197.5 & 9630.084 & 0.008 & 0.011 \\
152.5 & 9627.134 & 0.006 & 0.032 & 198.5 & 9630.150 & 0.004 & 0.017 \\
153.5 & 9627.200 & 0.006 & 0.039 & 199.5 & 9630.216 & 0.004 & 0.023 \\
154.5 & 9627.265 & 0.004 & 0.030 & 200.5 & 9630.281 & 0.004 & 0.014 \\
155.5 & 9627.331 & 0.008 & 0.036 & 201.5 & 9630.347 & 0.004 & 0.021 \\
156.5 & 9627.396 & 0.006 & 0.027 & 202.5 & 9630.413 & 0.004 & 0.027 \\
157.5 & 9627.462 & 0.004 & 0.033 & 203.5 & 9630.479 & 0.006 & 0.033 \\
158.5 & 9627.527 & 0.006 & 0.024 & 204.5 & 9630.545 & 0.008 & 0.039 \\
159.5 & 9627.593 & 0.008 & 0.031 & 205.5 & 9630.610 & 0.004 & 0.030 \\
160.5 & 9627.658 & 0.008 & 0.022 & 206.5 & 9630.676 & 0.004 & 0.037 \\
161.5 & 9627.724 & 0.010 & 0.028 & 207.5 & 9630.741 & 0.008 & 0.028 \\
162.5 & 9627.789 & 0.006 & 0.019 & 208.5 & 9630.806 & 0.006 & 0.019 \\
163.5 & 9627.854 & 0.008 & 0.010 & 209.5 & 9630.872 & 0.006 & 0.025 \\
164.5 & 9627.920 & 0.006 & 0.016 & 210.5 & 9630.937 & 0.006 & 0.016 \\
165.5 & 9627.986 & 0.006 & 0.023 & 211.5 & 9631.002 & 0.008 & 0.007 \\
166.5 & 9628.051 & 0.008 & 0.014 & 212.5 & 9631.067 & 0.006 & -0.002 \\
167.5 & 9628.117 & 0.006 & 0.020 & 214.5 & 9631.198 & 0.010 & -0.004 \\
168.5 & 9628.182 & 0.004 & 0.011 & 215.5 & 9631.264 & 0.006 & 0.002 \\
169.5 & 9628.248 & 0.004 & 0.017 & 217.5 & 9631.394 & 0.008 & -0.016 \\
170.5 & 9628.313 & 0.006 & 0.008 & 218.5 & 9631.460 & 0.010 & -0.010 \\
171.5 & 9628.378 & 0.004 & 0.000 & 220.5 & 9631.589 & 0.006 & -0.043 \\
172.5 & 9628.445 & 0.004 & 0.021 & 222.5 & 9631.722 & 0.006 & -0.015 \\
173.5 & 9628.510 & 0.004 & 0.012 & 225.5 & 9631.917 & 0.004 & -0.042 \\
174.5 & 9628.575 & 0.006 & 0.003 & 226.5 & 9631.981 & 0.004 & -0.066 \\
175.5 & 9628.641 & 0.004 & 0.009 & 227.5 & 9632.046 & 0.004 & -0.075 \\
176.5 & 9628.706 & 0.006 & 0.000 & 229.5 & 9632.180 & 0.004 & -0.032 \\
177.5 & 9628.772 & 0.006 & 0.007 & 230.5 & 9632.248 & 0.004 & 0.005 \\
178.5 & 9628.838 & 0.008 & 0.013 & 231.5 & 9632.314 & 0.004 & 0.011 \\
179.5 & 9628.904 & 0.008 & 0.019 & 232.5 & 9632.377 & 0.004 & -0.028 \\
180.5 & 9628.970 & 0.006 & 0.026 & 233.5 & 9632.436 & 0.004 & -0.129 \\
181.5 & 9629.035 & 0.006 & 0.017 & 234.5 & 9632.505 & 0.004 & -0.077 \\
182.5 & 9629.101 & 0.008 & 0.023 & 235.5 & 9632.575 & 0.004 & -0.010 \\
183.5 & 9629.166 & 0.004 & 0.014 & 237.5 & 9632.705 & 0.006 & -0.027 \\
184.5 & 9629.231 & 0.006 & 0.005 & 239.5 & 9632.830 & 0.004 & -0.122 \\
\hline
\end{tabular}
 \end{small}
\end{table*}  


\begin{table*}[!ht]
\renewcommand\thetable{9. Part I}
 \centering
 \begin{small}
  \caption{Times of maxima for ordinary superhumps in the light curve of ER UMa in the 2024 February superoutburst.}
\smallskip
  \begin{tabular}{@{}|l|c|l|r|l|c|l|r|@{}}
  \hline
$E$       & Times of maxima  &  Error  &  $O-C$ &  $E$         & Times of maxima  &  Error  &  $O-C$ \\
         &BJD$_{\text{max}} - 2450000$    &   &  [cycles] &   & BJD$_{\text{max}}-2450000$   &    &  [cycles] \\

\hline
0 & 10351.522 & 0.004 & -0.079 & 27 & 10353.298 & 0.004 & 0.003 \\
1 & 10351.594 & 0.004 & 0.018 & 28 & 10353.363 & 0.006 & -0.006 \\
2 & 10351.660 & 0.004 & 0.025 & 29 & 10353.428 & 0.006 & -0.015 \\
3 & 10351.727 & 0.004 & 0.047 & 30 & 10353.495 & 0.006 & 0.007 \\
4 & 10351.792 & 0.004 & 0.038 & 35 & 10353.821 & 0.004 & -0.022 \\
5 & 10351.857 & 0.004 & 0.029 & 36 & 10353.888 & 0.006 & -0.000 \\
6 & 10351.921 & 0.004 & 0.005 & 37 & 10353.954 & 0.006 & 0.006 \\
7 & 10351.987 & 0.004 & 0.011 & 38 & 10354.018 & 0.008 & -0.018 \\
8 & 10352.054 & 0.004 & 0.033 & 39 & 10354.085 & 0.008 & 0.004 \\
9 & 10352.118 & 0.004 & 0.009 & 40 & 10354.151 & 0.004 & 0.010 \\
10 & 10352.185 & 0.004 & 0.031 & 41 & 10354.215 & 0.004 & -0.014 \\
11 & 10352.250 & 0.002 & 0.022 & 42 & 10354.282 & 0.004 & 0.008 \\
12 & 10352.316 & 0.002 & 0.028 & 43 & 10354.346 & 0.004 & -0.016 \\
13 & 10352.381 & 0.002 & 0.019 & 44 & 10354.411 & 0.004 & -0.025 \\
14 & 10352.447 & 0.004 & 0.026 & 45 & 10354.476 & 0.004 & -0.034 \\
15 & 10352.513 & 0.004 & 0.032 & 46 & 10354.542 & 0.008 & -0.028 \\
16 & 10352.578 & 0.004 & 0.023 & 47 & 10354.606 & 0.004 & -0.052 \\
17 & 10352.643 & 0.004 & 0.015 & 48 & 10354.672 & 0.006 & -0.045 \\
18 & 10352.710 & 0.004 & 0.036 & 49 & 10354.739 & 0.006 & -0.024 \\
19 & 10352.774 & 0.004 & 0.012 & 51 & 10354.871 & 0.006 & -0.011 \\
20 & 10352.842 & 0.004 & 0.049 & 52 & 10354.934 & 0.008 & -0.050 \\
21 & 10352.907 & 0.006 & 0.040 & 53 & 10355.000 & 0.010 & -0.044 \\
22 & 10352.970 & 0.006 & 0.001 & 54 & 10355.066 & 0.010 & -0.037 \\
23 & 10353.038 & 0.006 & 0.038 & 55 & 10355.132 & 0.010 & -0.031 \\
24 & 10353.102 & 0.004 & 0.014 & 57 & 10355.262 & 0.008 & -0.048 \\
25 & 10353.170 & 0.004 & 0.051 & 58 & 10355.324 & 0.004 & -0.103 \\
26 & 10353.232 & 0.006 & -0.004 & 59 & 10355.392 & 0.006 & -0.066 \\       
\hline
\end{tabular}
 \end{small}
\end{table*}


\begin{table*}[!ht]
\renewcommand\thetable{9. Part II}
 \centering
 \begin{small}
  \caption{Times of maxima for late superhumps in the light curve of ER UMa in the 2024 February superoutburst.}
\smallskip
  \begin{tabular}{@{}|l|c|l|r|l|c|l|r|@{}}
  \hline
$E$       & Times of maxima  &  Error  &  $O-C$ &  $E$         & Times of maxima  &  Error  &  $O-C$ \\
         &BJD$_{\text{max}} - 2450000$    &   &  [cycles] &   & BJD$_{\text{max}}-2450000$   &    &  [cycles] \\

\hline
36.5 & 10353.918 & 0.004 & -0.043 & 181.5 & 10363.435 & 0.006 & 0.082 \\
37.5 & 10353.985 & 0.004 & -0.021 & 182.5 & 10363.499 & 0.006 & 0.058 \\
38.5 & 10354.051 & 0.006 & -0.015 & 183.5 & 10363.564 & 0.006 & 0.049 \\
39.5 & 10354.114 & 0.004 & -0.054 & 184.5 & 10363.630 & 0.006 & 0.055 \\
40.5 & 10354.183 & 0.004 & -0.002 & 185.5 & 10363.695 & 0.004 & 0.046 \\
41.5 & 10354.249 & 0.004 & 0.004 & 186.5 & 10363.761 & 0.006 & 0.053 \\
42.5 & 10354.310 & 0.006 & -0.065 & 187.5 & 10363.825 & 0.008 & 0.029 \\
43.5 & 10354.380 & 0.008 & 0.002 & 188.5 & 10363.890 & 0.008 & 0.020 \\
44.5 & 10354.446 & 0.006 & 0.009 & 190.5 & 10364.022 & 0.004 & 0.033 \\
45.5 & 10354.511 & 0.004 & 0.000 & 193.5 & 10364.214 & 0.004 & -0.039 \\
46.5 & 10354.575 & 0.004 & -0.024 & 195.5 & 10364.347 & 0.004 & -0.011 \\
47.5 & 10354.640 & 0.004 & -0.033 & 196.5 & 10364.416 & 0.008 & 0.041 \\
48.5 & 10354.705 & 0.004 & -0.042 & 197.5 & 10364.485 & 0.006 & 0.093 \\
49.5 & 10354.772 & 0.004 & -0.020 & 198.5 & 10364.550 & 0.006 & 0.084 \\
50.5 & 10354.836 & 0.004 & -0.044 & 199.5 & 10364.613 & 0.004 & 0.045 \\
51.5 & 10354.902 & 0.004 & -0.038 & 200.5 & 10364.677 & 0.006 & 0.021 \\
52.5 & 10354.966 & 0.006 & -0.062 & 201.5 & 10364.740 & 0.006 & -0.018 \\
53.5 & 10355.032 & 0.006 & -0.056 & 202.5 & 10364.806 & 0.006 & -0.012 \\
54.5 & 10355.098 & 0.004 & -0.049 & 203.5 & 10364.873 & 0.008 & 0.010 \\
55.5 & 10355.161 & 0.004 & -0.088 & 204.5 & 10364.935 & 0.004 & -0.045 \\
56.5 & 10355.225 & 0.008 & -0.113 & 205.5 & 10364.999 & 0.008 & -0.069 \\
57.5 & 10355.294 & 0.010 & -0.060 & 207.5 & 10365.130 & 0.004 & -0.071 \\
58.5 & 10355.356 & 0.010 & -0.115 & 208.5 & 10365.191 & 0.004 & -0.141 \\
59.5 & 10355.420 & 0.006 & -0.139 & 209.5 & 10365.268 & 0.004 & 0.033 \\
162.5 & 10362.185 & 0.004 & 0.020 & 210.5 & 10365.333 & 0.006 & 0.024 \\
163.5 & 10362.250 & 0.004 & 0.012 & 211.5 & 10365.400 & 0.008 & 0.046 \\
164.5 & 10362.316 & 0.004 & 0.018 & 213.5 & 10365.528 & 0.004 & -0.002 \\
165.5 & 10362.382 & 0.004 & 0.024 & 214.5 & 10365.592 & 0.008 & -0.026 \\
166.5 & 10362.447 & 0.008 & 0.016 & 215.5 & 10365.658 & 0.008 & -0.020 \\
167.5 & 10362.515 & 0.004 & 0.053 & 216.5 & 10365.725 & 0.008 & 0.002 \\
168.5 & 10362.579 & 0.006 & 0.028 & 217.5 & 10365.788 & 0.006 & -0.038 \\
169.5 & 10362.643 & 0.004 & 0.004 & 218.5 & 10365.859 & 0.006 & 0.045 \\
171.5 & 10362.774 & 0.004 & 0.002 & 219.5 & 10365.924 & 0.008 & 0.036 \\
172.5 & 10362.842 & 0.004 & 0.039 & 220.5 & 10365.987 & 0.008 & -0.003 \\
173.5 & 10362.906 & 0.008 & 0.015 & 221.5 & 10366.053 & 0.006 & 0.003 \\
174.5 & 10362.972 & 0.006 & 0.021 & 223.5 & 10366.178 & 0.006 & -0.090 \\
175.5 & 10363.039 & 0.004 & 0.043 & 224.5 & 10366.251 & 0.006 & 0.023 \\
176.5 & 10363.102 & 0.006 & 0.004 & 226.5 & 10366.379 & 0.006 & -0.025 \\
177.5 & 10363.168 & 0.004 & 0.010 & 227.5 & 10366.443 & 0.006 & -0.049 \\
178.5 & 10363.232 & 0.004 & -0.014 & 228.5 & 10366.508 & 0.006 & -0.058 \\
179.5 & 10363.297 & 0.008 & -0.023 & 230.5 & 10366.638 & 0.008 & -0.076 \\
180.5 & 10363.369 & 0.004 & 0.075 &  &  &  &  \\          
\hline
\end{tabular}
 \end{small}
\end{table*}


\end{document}